\documentclass[prd,aps,floats,twocolumn,floatfix,showpacs]{revtex4}
\usepackage[dvips]{epsfig}

\begin{document}

\def\sqr#1#2{{\vcenter{\hrule height.#2pt
   \hbox{\vrule width.#2pt height#1pt \kern#1pt
      \vrule width.#2pt}
   \hrule height.#2pt}}}
\def\square{{\mathchoice\sqr64\sqr64\sqr{3.0}3\sqr{3.0}3}}

\newcommand{\cjcell}[1]{\multicolumn{1}{c}{#1}}
\newcommand{\rjcell}[1]{\multicolumn{1}{r}{#1}}
\newcommand{\ljcell}[1]{\multicolumn{1}{l}{#1}}

\title{Deconfinement, gradient and cooling scales for pure SU(2) 
lattice gauge theory}

\author{Bernd A.\ Berg and David A.\ Clarke}

\affiliation{Department of Physics, Florida State University, 
             Tallahassee, FL 32306-4350, USA} 

 \date{\today } 

\begin{abstract}
We investigate the approach of pure SU(2) lattice gauge theory with 
the Wilson action to its continuum limit using the deconfining phase 
transition, the gradient flow and the cooling flow to set the scale. 
For the gradient and cooling scales we explore three different energy 
observables and two distinct reference values for the flow time. When 
the aim is to follow scaling towards the continuum limit, one gains at 
least a factor of 100 in computational efficiency by relying on the 
gradient instead of the deconfinement scale. Using cooling instead 
of the gradient flow one gains another factor of at least 34 in 
computational efficiency on the gradient flow part without any 
significant loss in the accuracy of scale setting. Concerning our 
observables, the message is to keep it simple. The Wilson action 
itself performs as well as or even better than the other two observables 
explored. Two distinct fitting forms for scaling are compared of which 
one connects to asymptotic scaling. Differences of the obtained 
estimates show that systematic errors of length ratios, though only 
about 1\%, can be considerably larger than statistical errors of
the same observables.
\end{abstract} 
\maketitle

\section{Introduction} \label{sec_intro}

We consider pure SU(2) lattice gauge theory (LGT) with the Wilson action
\begin{eqnarray} \label{S}
  &~& S\ =\ \beta\,\sum_{n,\mu<\nu}\left(1-\frac{1}{2}\,{\rm Tr}\,
  U^{\,\square}_{\mu\nu}(n)\right)\,,~~\beta=4/g_0^2\,,\qquad \\ &~&
  U^{\,\square}_{\mu\nu}(n)\ =\ U_{\mu}(n)\,U_{\nu}(n+\hat{\mu})\,
  U^{\dagger}_{\mu}(n+\hat{\nu})\, U^{\dagger}_{\nu}(n)\,. \qquad
\end{eqnarray}
Here $\hat{\mu}$, $\hat{\nu}$ are unit vectors in positive $\mu,\nu= 
1,2,3,4$ directions and $U^{\,\square}_{\mu\nu}$ is the product of SU(2) 
link variables along the boundary of a plaquette with one corner at site 
$n=(n_1,n_2,n_3,n_4)$ and $g_0$ is the bare coupling.

Due to its computational simplicity, pure SU(2) LGT is well suited as 
a showcase for computational methodology. Computational pitfalls or 
shortcomings are more easily identifiable than in more  complex systems 
like QCD. Furthermore, with modest CPU time resources, pure SU(2) LGT 
allows one to study the approach to the continuum limit for an entire 
range of suitable coupling constant values and lattice sizes. We 
investigate the approach of SU(2) LGT to its continuum limit using 
three different methods to set the scale:

\begin{enumerate}

\item The deconfining phase transition \cite{SY}. The deconfinement 
length scale is set by the inverse transition temperature times the
lattice spacing $a$. It has no ambiguities in its definition, but 
one needs to fit a number of parameters. Calculations of transition 
temperatures become very CPU time demanding with increasing lattice 
size.

\item L\"uscher's gradient flow \cite{L10}. When defining the gradient
scale one encounters a number of ambiguities. Once they are fixed, 
there are no parameters to fit. In our calculations the CPU time demands 
are reduced by at least two orders of magnitude when compared with the 
deconfinement scale.

\item Bonati and D'Elia \cite{BD14} noted that similar results as with 
the gradient scale are even more efficiently obtained using cooling 
\cite{B81} instead of the gradient flow. We demonstrate here in
quantitative detail that the cooling and gradient scales are for 
practical purposes equivalent. One gains another factor of at least
34 in computational efficiency on the gradient flow part by using 
cooling instead.

\end{enumerate}

Our results are obtained by Markov chain Monte Carlo (MCMC) simulations
for which we report the statistics in units of Monte Carlo plus 
Overrelaxation (MCOR) sweeps. One MCOR sweep updates each link once in 
a systematic order \cite{B04} with the Fabricius-Haan-Kennedy-Pendleton 
\cite{FHKP} heatbath algorithm and, in the same systematic order, twice 
by overrelaxation \cite{A88}. Using checkerboard coding \cite{BM82} 
and MPI Fortran, parallel updating of sublattices is implemented, and 
our SU(2) code is a scaled down version of the SU(3) code documented 
in Ref.~\cite{BW12}.

In the next section our estimates for the SU(2) deconfining phase 
transition are reported. Section~\ref{sec_grad} presents our results
for six SU(2) gradient scales. In section~\ref{sec_cool} the gradient 
flow is replaced by cooling. We analyze scaling and asymptotic scaling 
in section~\ref{sec_scaling}. Summary and conclusions are given in 
the final section~\ref{sec_sum}.

\section{Deconfinement scale} \label{sec_Tc}

We perform MCMC simulations on $N_s^3 N_{\tau}$ lattices and estimate
critical coupling constants $\beta_c(N_{\tau})$ up to $N_{\tau}=12$ 
by three-parameter fits 
\begin{eqnarray} \label{bc}
  \beta_c(N_s,N_{\tau})\ =\ \beta_c(N_{\tau}) 
  + a_1(N_{\tau})\,N_s^{a_2(N_{\tau})}
\end{eqnarray}
of pseudocritical $\beta_c(N_s,N_{\tau})$ values, where the fit 
parameters $\beta_c(N_{\tau})$ estimate the infinite volume values 
$\beta_c(N_s,\infty)$. Inverting the results of these fits defines 
the deconfining length scale
\begin{eqnarray} \label{Ntau}
  N_{\tau}(\beta_c)
\end{eqnarray}
to which we attach error bars by means of the equation
\begin{eqnarray} \label{dNt}
  \triangle N_\tau\ =\ \frac{N_\tau}{L_{10}^{1,3}(\beta_c)}\,\left[
  L_{10}^{1,3}(\beta_c)+L_{10}^{1,3}(\beta_c-\triangle\beta_c)\right]\,,
\end{eqnarray}
where the length scale $L_{10}^{1,3}(\beta)$ is introduced later in the
paper ($N_{\tau}$ error bars depend only mildly on the choice of the 
interpolation of its scaling behavior).

We use the locations of maxima of the Polyakov susceptibility to 
define pseudocritical $\beta_c(N_s,N_{\tau})$ values. Polyakov loops 
$P_{\vec{x}}$ are products of SU(2) matrices along straight lines 
in the $N_{\tau}$ direction. The argument $\vec{x}$ labels their 
locations on the spatial $N_s^3$ sublattice. From the sum over 
all Polyakov loops $P=\sum_{\vec{x}}P_{\vec{x}}$ one finds the 
susceptibility
\begin{equation} \label{chi}
  \chi(\beta)\ =\ \frac{1}{N_s^3} \left[ \langle P^2\rangle 
                      - \langle|P|\rangle^2\ \right]\,,
\end{equation} 
which is the analogue to the magnetic susceptibility of a spin system 
in three dimensions. We also implemented measurements of the thermal 
Polyakov loop susceptibility 
\begin{equation} \label{chi_beta}
  \chi_{~\atop T}(\beta)\ =\ \frac{1}{N_s^3} \frac{d~}{d\beta}\,
  \langle|P|\rangle\,,
\end{equation} 
but maxima are less pronounced than for $\chi(\beta)$.

\begin{table}[th] \centering 
\caption{\label{tab_Tc1} Pseudocritical $\beta$ values $N_s$: $\beta_c$. 
Error bars of $\beta_c$ are in parentheses.}
\begin{tabular}{|c|c|c|} \hline
$N_\tau=4$ &$N_\tau=6$ &$N_\tau=8$ \\ \hline
08: 2.30859~~(53) &12: 2.43900~~(33) &16: 2.52960~~(90)\\
12: 2.30334~~(33) &18: 2.43096~~(43) &24: 2.51678~~(43)\\
16: 2.30161~~(30) &20: 2.42973~~(11) &32: 2.51296~~(20)\\
20: 2.30085~~(17) &24: 2.42873~~(35) &40: 2.51192~~(12)\\
24: 2.30060~~(16) &28: 2.427939 (74) &44: 2.51150~~(11)\\
28: 2.30025~~(19) &30: 2.427690 (87) &48: 2.51119~~(11)\\
32: 2.299754 (99) &36: 2.427274 (67) &52: 2.51130~~(11)\\
40: 2.299593 (74) &44: 2.426827 (67) &56: 2.511096 (85)\\
48: 2.299452 (83) &48: 2.426756 (64) &64: 2.510635 (83)\\
56: 2.299435 (29) &56: 2.426605 (62) &72: 2.510716 (72)\\
                  &60: 2.426596 (55) &80: 2.510517 (79)\\ \hline
$\infty$: 2.299188 (61)&$\infty$: 2.426366 (52)&$\infty$: 2.510363 (71)
\\ $q= 0.56$         & $q=0.73$        & $q=0.14$ \\ \hline
$N_\tau=4\pm 0.00063$&$N_\tau=6\pm 0.0011$&$N_\tau=8\pm 0.0019$\\
\hline
\end{tabular} \end{table}

\begin{table}[th] \centering 
\caption{\label{tab_Tc2} Pseudocritical $\beta$ values $N_s$: 
        $\beta_c$ (continuation).} 
\begin{tabular}{|c|c|} \hline
$N_\tau=10$ &$N_\tau=12$ \\ \hline
20: 2.59961 (52) &                \\
24: 2.58909 (49) &24: 2.66317 (91)\\
28: 2.58497 (26) &\\
32: 2.58270 (27) &32: 2.64450 (39)\\
36: 2.58117 (13) &36: 2.64223 (33)\\
40: 2.58046 (26) &40: 2.64039 (26)\\
44: 2.58002 (17) &44: 2.63925 (24)\\
48: 2.57941 (15) &48: 2.63839 (27)\\
52: 2.57949 (23) &52: 2.63744 (19)\\
56: 2.57876 (18) &                \\
64: 2.57851 (15) &                \\ \hline
$\infty$: 2.57826 (14) &$\infty$: 2.63625 (35)\\
          $q=0.29$    &          $q=0.06$ \\ \hline
$N_\tau=10\pm 0.0045$&$N_\tau=12\pm 0.013$\\ \hline
\end{tabular} \end{table}

We use reweighting in small neighborhoods of the simulation points 
to extract pseudocritical $\beta$ values from the locations of the
maxima. The error bars are then estimated by repeating the entire
procedure for $\ge 32$ jackknife bins (see, e.g., \cite{B04}). Notably,
the estimates of pseudocritical $\beta$ values from the maxima of
(\ref{chi}) and (\ref{chi_beta}) may not fall into one reweighting 
range, though they have ultimately identical $N_s\to\infty$ limits. 
So, to reduce computational requirements one is pressed to make 
a decision in favor of one of them.

Together with their goodness of fit $q$ (for the definition see, e.g., 
Ref.~\cite{B04}), our pseudocritical $\beta_c$ estimates are compiled 
in Tables~\ref{tab_Tc1} and~\ref{tab_Tc2}. In previous literature Engels 
et al.~\cite{E96} studied $N_{\tau}=4$ extensively and demonstrated 
that it falls into the 3D Ising universality class. Their $N_s\to
\infty$ estimate $\beta_c(4)=2.29895$ (10) is marginally 
smaller than our estimate in Table~\ref{tab_Tc1} with $q=0.042$ from 
a Gaussian difference test (see, e.g., \cite{B04}). For $N_{\tau}$ 
values up to $N_{\tau}=8$ we found estimates in a paper by Lucini et 
al.~\cite{LTW04}. Gaussian difference tests with our estimates give 
$q=0.33$ and $q=0.67$ for $N_{\tau} = 4$ and~6, respectively. For 
$N_{\tau}=8$ their estimate $\beta_c(8)=2.5090$ (6) is somewhat 
lower than ours of Table~\ref{tab_Tc2}, which has an almost ten 
times smaller error bar than theirs. The Gaussian difference test 
gives $q=0.022$.

For $N_{\tau}=10$ and~12 calculations of the pseudocritical $\beta$ 
values from maxima of the Polyakov loop susceptibility become very 
CPU time consuming. The largest statistics we assembled consists of
slightly more than $2^{25}$ MCOR sweeps for the $40^3 12$ lattice. 
On even larger $N_{\tau}=10$ and 12 lattices we spent $2^{23}$ MCOR 
sweeps. The largest amounts of CPU time were not spent on the largest 
lattices because we were mainly feeding on the NERSC scavenger queue. 
For comparison, at $\beta=2.67$ we spent only $2^{19}$ MCOR sweeps 
on generating the $40^4$ lattice used for the gradient flow. Taking 
achieved error bars, lattice sizes and numbers of lattices needed in 
account, this amounts to improvements by factors of at least 100.
In view of the degrading of the deconfinement transition estimates 
with increasing lattice size, we also tried improved estimators 
\cite{P83}, performing the SU(2) integration explicitly. However, 
correlations between Polyakov loops 
turned out to be too strong to allow for major gains.

\begin{figure}[th]\begin{center} %
\epsfig{figure=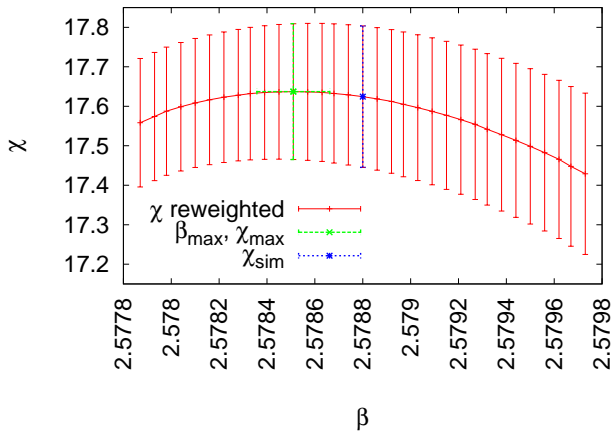,width=\columnwidth} 
\caption{Reweighting of the Polyakov loop susceptibility on our
$64^310$ lattice. \label{fig_CPsim}} 
\end{center} \vspace{-4mm} \end{figure} 

\begin{figure}[th]\begin{center} %
\epsfig{figure=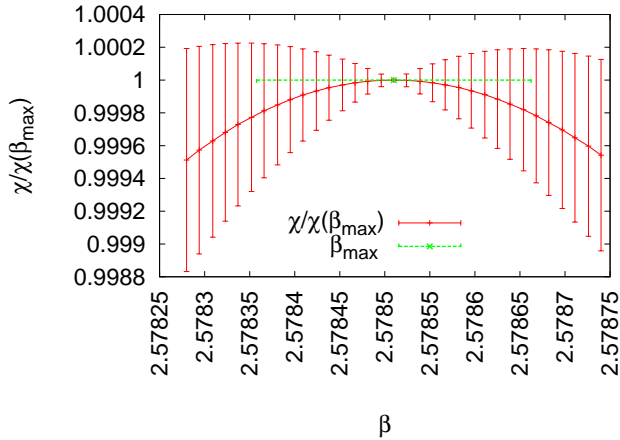,width=\columnwidth} 
\caption{Ratios of the Polyakov loop susceptibilities around
the $\beta_{\max}$ value of our previous figure. \label{fig_CPrat}} 
\end{center} \vspace{-4mm} \end{figure} 

For $N_{\tau}=10$ and 12 the reweighting curve about the simulation 
point $\beta_{\rm sim}$ becomes rather flat within large error bars. See 
Fig.~\ref{fig_CPsim} for an example. Therefore, one may be amazed about 
the astonishingly accurate estimate of the maximum position $\beta_{\max
}$. This is explained by the fact that all these error bars are strongly 
correlated, because they rely on reweighting of the same simulation. 
Dividing out the maximum value $\chi(\beta_{max})$ of the susceptibility 
in each jackknife bin, one is led to Fig.~\ref{fig_CPrat}, which 
projects out the central part around the maximum of the previous figure 
and makes the (jackknife) error bars of the $\beta_{\max}$ estimate 
plausible.

The scaling analysis of the $N_{\tau}(\beta_c)$ estimates 
of Tables~\ref{tab_Tc1} and~\ref{tab_Tc2} is performed in 
section~\ref{sec_scaling}.

\section{Gradient scale} \label{sec_grad}

Before coming to our central issue of scale setting we define the
SU(2) gradient flow, the observables used and sketch our generation
of MCMC data.

\subsection{Gradient flow}

With initial condition $U_{\mu}(n,0)=U_{\mu}(n)$ the gradient flow is 
defined \cite{L10} by the evolution equation
\begin{eqnarray} \label{dotU}
  \dot{U}_{\mu}(n,t)&=&-g_0^2\,
  \{\partial_{n,\mu}S[U(t)]\}\,U_{\mu}(n,t)\,.
\end{eqnarray}
Here the SU(2) link derivatives are given by
\begin{eqnarray} \label{partial}
  \partial_{n,\mu}f(U) = i\sum_{j=1}^3\sigma_j\left.\frac{d~}{ds}
  f(e^{isX^j}U)\right|_{s=0}\,,
\end{eqnarray}
where $\sigma_j$ are the Pauli matrices and
\begin{eqnarray} \label{Xjyx}
  X^j(m,\nu)\ =\ \cases{\sigma^j\ {\rm if}\ (m,\nu)=(n,\mu)\,,\cr 
                        0~~{\rm otherwise\,.} }
\end{eqnarray}
We use the notation $U_{\mu}^{\,\square}$ for the sum of plaquette 
matrices containing the link matrix $U_{\mu}$. With the definition 
of the staple matrix, 
\begin{eqnarray} \label{Us} 
  U^{\sqcup}_{\mu}(n) &=& \sum_{\nu\ne\mu} U_{\nu}(n)\,
  U_{\mu}(n+\hat{\nu})\,U^{\dagger}_{\nu}(n+\hat{\mu})\\ \nonumber
  &+& \sum_{\nu\ne\mu} U^{\dagger}_{\nu}(n-\hat{\nu})\,
  U_{\mu}(n-\hat{\nu})\,U_{\mu}(n-\hat{\nu}+\hat{\mu})\,,
\end{eqnarray}
this is
\begin{eqnarray} \label{U}
  U_{\mu}^{\,\square}(n) = U_{\mu}(n)\,U^{\sqcup}_{\mu}(n)^{\dagger}\,.
\end{eqnarray}
For the SU(2) link derivative (\ref{partial}) one finds the simple 
equation
\begin{eqnarray} \label{partialS}
  g_0^2\,\partial_{n,\mu}S(U)\ =\ \frac{1}{2} \left(
  U_{\mu}^{\,\square}(n)-U_{\mu}^{\,\square}(n)^{\dagger}\right)\,,
\end{eqnarray}
and we calculate the time evolution (\ref{dotU}) using the Runge-Kutta 
scheme described in appendix~C of \cite{L10} with
\begin{eqnarray} \label{Z}
  Z_i = \epsilon\,Z(W_i)\,,~~Z(W_i)= \frac{1}{2}\,\left(W_i - 
  W_i^{\dagger}\right)\,,
\end{eqnarray}
$W_0=U_{\mu}(n,t)$ as starting values and $\epsilon=0.01$.

\subsection{Observables}

For the lattice expectation values of the time dependent plaquette 
matrices we use the parametrization
\begin{eqnarray} \label{ai}
  \left<U^{\,\square}(t)\right>_L\ =\ a_0(t)\,\sigma_0 
               + i\,\sum_{i=1}^3a_i(t)\,\sigma_i\,,
\end{eqnarray}
where we suppress the $\mu\nu$ subscripts and $\sigma_0$ is the 
$2\times 2$ unit matrix, supplementing the Pauli matrices $\sigma_j$. 
As observables we use three definitions of the energy density: 
$E_0(t)$, $E_1(t)$ and $E_4(t)$. Up to a constant factor
\begin{eqnarray} \label{E0}
  E_0(t)\ =\ 2\,\left[1-a_0(t)\right]\,
\end{eqnarray}
is the usual Wilson action, i.e., becomes $\sim F_{\alpha\beta}
F_{\alpha\beta}$ in the continuum limit. The definition
\begin{eqnarray} \label{E1}
  E_1(t)\ =\ \sum_{i=1}^3 a_i(t)^2
\end{eqnarray}
has the same continuum limit as $E_0$. Finally, we denote by $E_4(t)$
L\"uscher's \cite{L10} energy density which averages over the four 
plaquettes attached to each site $n$ in a fixed $\mu\nu$, $\mu\ne\nu$ 
plane, i.e.,
\begin{eqnarray} \label{E4}
  E_4(t) &=& \sum_{i=1}^3 b_i(t)^2\,,\\ \nonumber 
  b_i(t) &=& \frac{1}{4}\,
  \left(a_i^{ul}+a_i^{ur}+a_i^{dl}+a_i^{dr}\right),
\end{eqnarray}
where the superscripts of $a_i$ stand for up ($u$), left ($l$), right 
($r$), and down ($d$) in a fixed $\mu\nu$ plane with respect to $n$ 
(drawn in Fig.~1 of \cite{L10}). The functions
\begin{eqnarray} \label{yi}
  y_i(t)\ =\ t^2\,E_i(t)\,,~~(i=0,1,4) 
\end{eqnarray}
are used to set the three gradient scales by choosing appropriate fixed 
values $y_i^0$ and iterating the time evolution (\ref{yi}) until
\begin{eqnarray} \label{t0i}
  y_i^0\ =\ (t_i^0)^2\,E_i(t_i^0)
\end{eqnarray}
is reached. As function of $\beta$ the observable
\begin{eqnarray} \label{s0i}
  s_i^0(\beta)\ =\ \sqrt{t_i^0(\beta)}
\end{eqnarray}
then scales like a length provided the following conditions are met:
\begin{enumerate}
\item Lattice sizes have to be chosen so that $N_{\min}\gg \sqrt{8}\,
s_i^0$ holds, where $\sqrt{8}\,s_i^0$ is the smoothing range \cite{L10} 
and $N_{\min}=\min\{N_i,i=1,2, 3,4\}$ for simulations on a $N_1N_2N_3
N_4$ lattice.
\item The values of $\beta$ have to be large enough to be in the 
SU(2) scaling region.
\item The values of $y_i^0$ have to be large enough so that $\sqrt{8}\,
s_i^0\gg 1$ holds for the smallest used flow time.
\end{enumerate}

\subsection{Data generation and analysis} \label{sec_dgrad}

Our numerical results rely on MCMC simulations for the $\beta$ values
and lattice sizes given in Table~\ref{tab_t01} and (identically) in
subsequent tables. In each run $128=2^7$ configurations were generated 
and on each of them the gradient flow was performed. To optimize our 
use of computational resources, we followed the rule of \cite{B04} and 
allocated our CPU time in approximately equal parts to generation 
of configurations and to measurements (gradient flow). Subsequent 
configurations were separated by $2^{11}$ to $2^{13}$ MCOR sweeps 
where the increase from $2^{11}$ to larger numbers of MCOR sweeps 
is essentially 
enforced by the number of gradient sweeps needed to reach the $y^0_i$ 
target values. The dividing line from $2^{11}$ to $2^{12}$ sweeps is 
between $\beta=2.574$ and $\beta=2.62$, and from $2^{12}$ to $2^{13}$ 
between $\beta=2.67$ and $\beta=2.71$.  We estimated integrated 
autocorrelation times $\tau_{\rm int}$ for the time series of 128 
measured scale values and found all $\tau_{\rm int}$ compatible with~1 
(in units of the number of sweeps between the configurations). So, 
the data are considered to be statistically independent. Error bars 
were calculated by the jackknife method with respect to the 128 
configurations. Mostly, we used $N^4$ lattices with the exception 
of $24^348$ and $32^364$, which mirror lattices used in \cite{L10}. 
The scale estimates from these asymmetric lattices are consistent with 
those we obtained from $N^4$ lattices. 

\subsection{Scale setting} \label{sec_scale}

\begin{table}[th] 
\centering 
\caption{\label{tab_t01}{Gradient length scale for its $y^{01}_i$ 
set (\ref{yi01}).}} \smallskip
\begin{tabular}{|c|c|c|c|c|} \hline
$\beta$&Lattice &$L_1=s_0^{01}$&$L_2=s_1^{01}$&$L_3=s_4^{01}$\\ \hline
2.3&   $8^4$  &1.361~~(13) &1.361~~(13) &1.359~~(15)\\ 
2.3&   $12^4$ &1.3538 (52) &1.3538 (50) &1.2955 (88)\\ 
2.3&   $16^4$ &1.3593 (28) &1.3589 (27) &1.2756 (75)\\ \hline 
2.43&  $12^4$ &2.126~~(20) &2.115~~(20) &2.038~~(20)\\ 
2.43&  $16^4$ &2.0961 (91) &2.0848 (90) &1.964~~(14)\\ 
2.43&  $24^4$ &2.1066 (41) &2.0952 (40) &1.974~~(11)\\ 
2.43&  $28^4$ &2.1023 (30) &2.0911 (30) &1.9666 (98)\\ \hline
2.51&  $16^4$ &2.730~~(21) &2.715~~(21) &2.603~~(23)\\ 
2.51&  $20^4$ &2.766~~(15) &2.750~~(15) &2.585~~(20)\\ 
2.51&  $28^4$ &2.7590 (73) &2.7428 (73) &2.570~~(14)\\ \hline
2.574& $20^4$ &3.389~~(26) &3.369~~(26) &3.166~~(28)\\ 
2.574& $24^4$ &3.395~~(17) &3.374~~(17) &3.175~~(22)\\ 
2.574& $32^4$ &3.406~~(11) &3.385~~(11) &3.193~~(17)\\ 
2.574& $40^4$ &3.4103 (72) &3.3896 (71) &3.149~~(16)\\ \hline
2.62&  $24^4$ &3.993~~(28) &3.968~~(28) &3.711~~(35)\\ 
2.62 &$24^348$&3.947~~(22) &3.923~~(21) &3.699~~(26)\\ 
2.62&  $28^4$ &3.950~~(20) &3.926~~(20) &3.704~~(24)\\ 
2.62&  $40^4$ &3.954~~(10) &3.9293 (99) &3.672~~(19)\\ \hline
2.67&  $28^4$ &4.680~~(33) &4.651~~(33) &4.350~~(39)\\ 
2.67&  $32^4$ &4.651~~(27) &4.622~~(27) &4.350~~(33)\\ 
2.67&  $40^4$ &4.622~~(17) &4.593~~(17) &4.297~~(24)\\ \hline
2.71&  $32^4$ &5.217~~(37) &5.185~~(37) &4.867~~(42)\\ 
2.71&  $36^4$ &5.252~~(33) &5.220~~(33) &4.852~~(42)\\ 
2.71&  $40^4$ &5.199~~(22) &5.167~~(22) &4.817~~(27)\\ \hline 
2.751&$32^364$&5.879~~(35) &5.843~~(34) &5.466~~(39)\\ 
2.751& $36^4$ &5.893~~(38) &5.856~~(38) &5.465~~(48)\\ 
2.751& $40^4$ &5.909~~(34) &5.872~~(34) &5.457~~(41)\\ \hline
2.816& $44^4$ &7.092~~(48) &7.049~~(47) &6.530~~(54)\\ \hline
2.875& $52^4$ & 8.510~~(64)& 8.456~~(65)& 7.883~~(68) \\ \hline
\end{tabular} \end{table} 

\begin{table}[th] 
\centering 
\caption{\label{tab_t02}{Gradient length scale for its $y^{02}_i$ 
set (\ref{yi02}).}} \smallskip
\begin{tabular}{|c|c|c|c|c|} \hline
$\beta$&Lattice&$L_4=s_0^{02}$&$L_5=s_1^{02}$&$L_6=s_4^{02}$\\ \hline
2.3&   $ 8^4$ &1.897~~(24) & 1.897~~(24) & 1.900 (25) \\ 
2.3&   $12^4$ &1.8905 (84) & 1.8897 (83) & 1.824 (12) \\ 
2.3&   $16^4$ &1.8963 (48) & 1.8956 (48) & 1.807 (11) \\ \hline 
2.43&  $12^4$ &2.849~~(34) & 2.842~~(33) & 2.771 (34) \\ 
2.43&  $16^4$ &2.791~~(15) & 2.784~~(15) & 2.653 (20) \\ 
2.43&  $24^4$ &2.8044 (66) & 2.7968 (65) & 2.644 (15) \\ 
2.43&  $28^4$ &2.7994 (48) & 2.7920 (47) & 2.645 (13) \\ \hline 
2.51&  $16^4$ &3.586~~(34) & 3.575~~(34) & 3.436 (34) \\ 
2.51&  $20^4$ &3.653~~(25) & 3.642~~(25) & 3.453 (29) \\ 
2.51&  $28^4$ &3.624~~(12) & 3.613~~(12) & 3.406 (19) \\ \hline
2.574& $20^4$ &4.437~~(39) & 4.423~~(39) & 4.178 (44) \\ 
2.574& $24^4$ &4.429~~(26) & 4.415~~(26) & 4.171 (29) \\ 
2.574& $32^4$ &4.454~~(15) & 4.440~~(15) & 4.219 (22) \\ 
2.574& $40^4$ &4.458~~(12) & 4.444~~(11) & 4.175 (21) \\ \hline
2.62&  $24^4$ &5.252~~(46) & 5.233~~(45) & 4.916 (49) \\ 
2.62& $24^348$&5.135~~(33) & 5.119~~(33) & 4.868 (38) \\ 
2.62&  $28^4$ &5.145~~(30) & 5.129~~(30) & 4.849 (32) \\ 
2.62&  $40^4$ &5.156~~(16) & 5.140~~(16) & 4.827 (26) \\ \hline
2.67&  $28^4$ &6.131~~(53) & 6.110~~(53) & 5.740 (60) \\ 
2.67&  $32^4$ &6.057~~(40) & 6.038~~(40) & 5.719 (46) \\ 
2.67&  $40^4$ &6.020~~(27) & 6.000~~(27) & 5.645 (32) \\ \hline
2.71&  $32^4$ &6.776~~(55) & 6.754~~(55) & 6.357 (56) \\ 
2.71&  $36^4$ &6.831~~(50) & 6.809~~(50) & 6.401 (57) \\ 
2.71&  $40^4$ &6.773~~(32) & 6.751~~(32) & 6.334 (39) \\ \hline 
2.751&$32^364$&7.642~~(51) & 7.617~~(51) & 7.179 (57) \\ 
2.751& $36^4$ &7.659~~(60) & 7.633~~(59) & 7.161 (68) \\ 
2.751& $40^4$ &7.694~~(50) & 7.668~~(50) & 7.211 (59) \\ \hline
\end{tabular} \end{table} 

\begin{figure}[th]\begin{center} 
\epsfig{figure=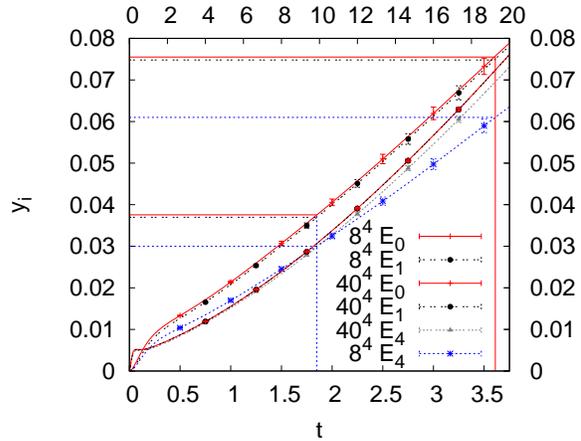,width=\columnwidth} 
\caption{Gradient flows $y_i(t)$ for the energy densities $E_0$, $E_1$ 
and $E_4$ at $\beta=2.3$ on an $8^4$ lattice ($t$ on lower abscissa)
and at $\beta=2.574$ on a $40^4$ lattice ($t$ on upper abscissa). The
up-down order in the legend agrees on the right-hand side with that of 
the curves.  \label{fig_t40}} 
\end{center} \vspace{-4mm} \end{figure} 

From estimates of the deconfinement $\beta_c(N_{\tau})$ values we know 
that it only makes sense to investigate SU(2) scaling for $\beta\ge 
2.29$, $N_{\tau}\ge 4$. The smallest $N_s^3\,4$ lattice size that can 
be used for the $N_{\tau}=4$, $N_s\to\infty$ finite size extrapolation 
is given by $N_s=8$. Therefore, it is natural to start our gradient flow 
simulations at $\beta=2.3$ on an $8^4$ lattice and to work from there 
up to larger $\beta$ values and lattice sizes. It is of interest to 
control scaling violations at the lower end of the scaling region,
because simulations there are less expensive than at larger $\beta$. 

The upper two curves (red, black online) and the, ultimately, lowest
(blue online) curve of Fig.~\ref{fig_t40} show $y_i(t)$, ($i=0,1,4$) 
from simulations on an $8^4$ lattice ($t$ on the lower abscissa). While 
the plots corresponding to $E_0$ and $E_1$ fall practically on top of 
one another, they deviate from the plot for $E_4$. This is due to finite
lattice size corrections as well as scaling corrections in $\beta$.
These corrections are much smaller for the other three curves which are
from simulations at $\beta=2.574$ on a $40^4$ lattice. The corresponding
$t$ values are on the upper abscissa and chosen so that the largest 
$y$-values reached agree approximately with those from the $8^4$ 
lattice.

\begin{figure}[th]\begin{center} 
\epsfig{figure=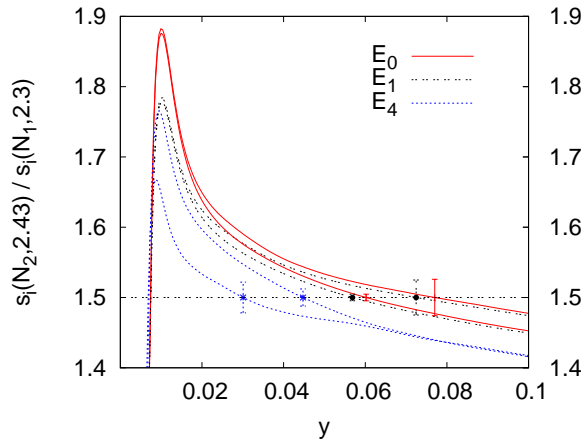,width=\columnwidth} 
\caption{Gradient flow ratios as functions of $y$. Outer curves:
$s_i(12,2.43)/s_i(8,2.3)$. Inner curves: $s_i(24,2.43)/s_i(16,2.3)$.
\label{fig_rat}} 
\end{center} \vspace{-4mm} \end{figure}

The question is this: How does one pick a set of $y_i^0$ values that 
defines suitable $s^0_i$ scales according to Eqs.~(\ref{t0i}) and 
(\ref{s0i})? To minimize CPU time, one likes to keep the lattice size 
and $s_0^i$ as small as possible. On the other hand, smaller $s_0^i$ 
values imply larger discretization (finite lattice spacing) corrections 
and too small lattices imply finite size corrections. It is at this 
point that one encounters considerable ambiguities in the definition 
of gradient (and similarly cooling) scales.

In our context it is natural to define $y_i^0$ values so that our 
initial estimates from the $s^0_i$ scales are consistent with those 
from low $\beta_c(N_{\tau}$) values. Lowest reasonable starting values 
for $\beta$, corresponding approximately to the $\beta_c(4)$ and 
$\beta_c(6)$ estimates of Table~\ref{tab_Tc1}, are $\beta_1=2.3$ and 
$\beta_2=2.43$. In Fig.~\ref{fig_rat} we plot ratio functions 
\begin{eqnarray} \label{ri}
  \frac{s_i(N_2,\beta_2=2.43)}{s_i(N_1,\beta_1=2.3)}\,(y)
\end{eqnarray}
for $(N_2,N_1)=(12,8)$ and $(24,16)$. On the 1.5 line the outer curves 
correspond to $(12,8)$ and the inner curves (using the same colors) to 
$(24,16)$. To prevent the figure from becoming too convoluted, error 
bars are only given on this line. As one expects from Fig.~\ref{fig_t40}, 
the $y$ values of the $E_4$ crossing points are apart from those of 
$E_0$ and $E_1$. The difference is considerably reduced when finite 
lattice size corrections are remediated by moving to (24,16) lattices. 
The remaining difference should mainly be attributed to corrections 
in $\beta$ (i.e., finite lattice spacing corrections).

One may have expected a plateau in the neighborhood of the 1.5 line, 
indicating that the ratios do not depend on the precise choice of the 
$y_i^0$ target values. Instead, without using the deconfinement result 
as input, another uncertainty in the choice of the $y_i^0$ target 
values would exist. 

In the following we use the outer values of 
Fig.~\ref{fig_rat} and explore whether their differences result in 
seriously distinct scaling behavior. Starting off with $\beta=2.3$, 
we are exploring two gradient scales:
\begin{enumerate}
\item We define the $y_i^{01}$ scale so that the $E_4$ observable
      delivers $s_4^{01}(12,2.43)/s_4^{01}(8,2.3)=1.5$.
\item We define the $y_i^{02}$ scale so that for the $E_i$, $i=0,1$,
      observables $s_i^{02}(12,2.43)/s_i^{02}(8,2.3)=1.5$ holds.
\end{enumerate}
For the first case we find $y_4^{01}=0.030$ from Fig.~\ref{fig_rat}.
Using $y_4(t)$ depicted in Fig.~\ref{fig_t40}, $y_4^{01}=0.030$ converts 
into the $t$ value $t^{01}=1.85$ for the flow time, as indicated by 
a vertical line. Its intersections with the $y_i(t)$ functions define 
our first set of $y_i^0$ target values 
\begin{eqnarray} \label{yi01}
   y^{01}_0=0.0376\,,~~~y^{01}_1=0.0370\,,~~~y^{01}_4=0.030\,.
\end{eqnarray}
Similarly, our second set of $y_i^0$ values is derived from $t^{02}=
3.61$, which is the average value of $t$ of the relevant crossing points 
of the $E_0$ and $E_1$ observables. This $t^{02}$ value is also shown 
as a vertical line in Fig.~\ref{fig_t40} and leads to
\begin{eqnarray} \label{yi02}
  y^{02}_0=0.0755\,,~~~y^{02}_1=0.0748\,,~~~y^{02}_4=0.061\,.
\end{eqnarray}
Length scale values 
\begin{eqnarray} \label{s0scale}
  s_i^{0j}(\beta)\ =\ \sqrt{t_i^{0j}(\beta)}\,,~~i=0,1,4,~~j=1,2
\end{eqnarray} 
are obtained when the gradient flow hits the corresponding $y_i^{0j}$ 
definitions of Eqs.~(\ref{yi01}) or (\ref{yi02}). Our MCMC estimates for 
them are reported in Tables~\ref{tab_t01} and~\ref{tab_t02}. For later 
convenience we label the length scales by $L_1$ to $L_6$ as defined in 
the first row of the tables. We are led to $\sqrt{8\,t^{01}}\approx 
3.85$ and $\sqrt{8\,t^{02}}\approx 5.37$ as our smallest values for 
the smoothing range. This is below and above the starting value $\sqrt
{8\,t^0}\approx 4.77$ of Ref.~\cite{L10} taken at $\beta=5.96$ in the 
SU(3) scaling region. Comparing the SU(3) deconfinement transition 
values $\beta_c$ for $N_{\tau}=4$, 6, 8 (see, e.g., Ref.~\cite{Fr15})
with those for SU(2) and performing interpolations of the $\beta_c$ 
values, this corresponds roughly to $\beta=2.46$ for SU(2), where our 
lower smoothing range has increased to at least 6.64. So, our lower 
smoothing range is also effectively larger than the one of~\cite{L10}. 

For each $\beta$ value several lattice sizes are listed in 
Tables~\ref{tab_t01} and~\ref{tab_t02} to control finite size 
corrections. In most cases they are sufficiently weak to be swallowed 
by the statistical error bars. Exceptions are the ${s}_4^{0j}$ estimates 
on $8^4$ and $12^4$ lattices at $\beta=2.3$ and 2.43, which appear to 
be too small to accommodate $E_4$. Up to $\beta=2.751$ lattices of size 
$40^4$ appear to be large enough so that finite size 
corrections can be neglected. Larger lattices would just increase 
statistics due to self-averaging. For our largest lattices with 
$\beta=2.816$ and 2.875 the gradient flow was designed too 
short to reach the $y_i^{02}$ targets~(\ref{yi02}). 

\section{Cooling scale} \label{sec_cool}

\begin{figure}[th]\begin{center} 
\epsfig{figure=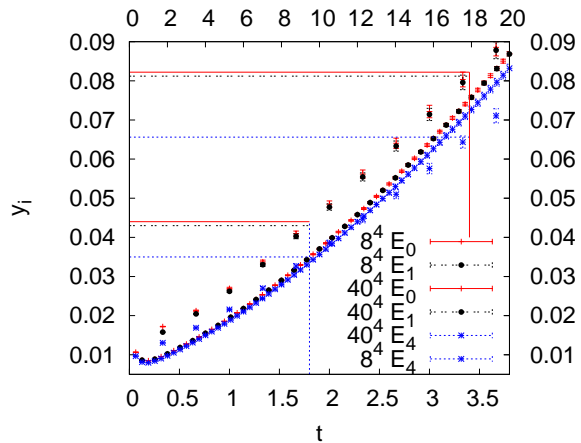,width=\columnwidth} 
\caption{Cooling flows $y_i(t)$ for the energy densities $E_0$, $E_1$ 
and $E_4$ at $\beta=2.3$ on an $8^4$ lattice ($t$ on lower abscissa)
and at $\beta=2.574$ on a $40^4$ lattice ($t$ on upper abscissa). 
\label{fig_t0cool}} \end{center} \end{figure}  

\begin{table}[th] 
\centering 
\caption{\label{tab_ct01}{Cooling length scale for its $y^{01}_i$ 
set (\ref{cyi01}).}} \smallskip
\begin{tabular}{|c|c|c|c|c|} \hline
$\beta$&Lattice&$L_7=s_0^{01}$&$L_8=s_1^{01}$&$L_9=s_4^{01}$\\ \hline
2.3  & $ 8^4$ & 1.342~~(12)& 1.337~~(12)& 1.342~~(14) \\ 
2.3  & $12^4$ & 1.3391( 47)& 1.3343 (45)& 1.2730 (85) \\ 
2.3  & $16^4$ & 1.3433 (24)& 1.3385 (23)& 1.2575 (74) \\ \hline
2.43 & $12^4$ & 2.111~~(19)& 2.092~~(18)& 2.013~~(20) \\ 
2.43 & $16^4$ & 2.0837 (90)& 2.0653 (90)& 1.951~~(13) \\ 
2.43 & $24^4$ & 2.0929 (38)& 2.0744 (38)& 1.947~~(11) \\ 
2.43 & $28^4$ & 2.0892 (28)& 2.0707 (28)& 1.9446 (95) \\ \hline
2.51 & $16^4$ & 2.728~~(19)& 2.703~~(19)& 2.587~~(23) \\ 
2.51 & $20^4$ & 2.753~~(14)& 2.727~~(14)& 2.567~~(20) \\ 
2.51 & $28^4$ & 2.7522 (68)& 2.7267 (66)& 2.548~~(15) \\ \hline
2.574& $20^4$ & 3.396~~(25)& 3.365~~(24)& 3.157~~(26) \\ 
2.574& $24^4$ & 3.389~~(16)& 3.357~~(16)& 3.155~~(22) \\ 
2.574& $32^4$ & 3.4001 (97)& 3.3686 (95)& 3.153~~(17) \\ 
2.574& $40^4$ & 3.4048 (69)& 3.3730 (67)& 3.137~~(17) \\ \hline
2.62 & $24^4$ & 3.988~~(26)& 3.949~~(26)& 3.717~~(32) \\ 
2.62 &$24^348$& 3.949~~(20)& 3.912~~(19)& 3.688~~(25) \\ 
2.62 & $28^4$ & 3.952~~(19)& 3.915~~(19)& 3.680~~(23) \\ 
2.62 & $40^4$ & 3.9509 (95)& 3.9137 (93)& 3.645~~(22) \\ \hline
2.67 & $28^4$ & 4.676~~(32)& 4.631~~(31)& 4.314~~(39) \\ 
2.67 & $32^4$ & 4.644~~(27)& 4.600~~(26)& 4.282~~(31) \\
2.67 & $40^4$ & 4.618~~(17)& 4.574~~(16)& 4.298~~(26) \\ \hline
2.71 & $32^4$ & 5.216~~(36)& 5.167~~(35)& 4.833~~(41) \\ 
2.71 & $36^4$ & 5.256~~(31)& 5.207~~(31)& 4.803~~(42) \\ 
2.71 & $40^4$ & 5.203~~(21)& 5.154~~(21)& 4.794~~(28) \\ \hline
2.751&$32^364$& 5.874~~(32)& 5.819~~(32)& 5.437~~(37) \\ 
2.751& $36^4$ & 5.892~~(36)& 5.836~~(35)& 5.478~~(49) \\ 
2.751& $40^4$ & 5.913~~(32)& 5.857~~(32)& 5.434~~(40) \\ \hline
2.816& $44^4$ & 7.105~~(45)& 7.039~~(45)& 6.511~~(55) \\ \hline
2.875& $52^4$ & 8.514~~(60)& 8.433~~(59)& 7.825~~(68) \\ \hline
\end{tabular} 
\end{table} 

\begin{table}[th] 
\centering 
\caption{\label{tab_ct02}{Cooling length scale for its $y^{02}_i$ 
set (\ref{cyi02}).}} \smallskip
\begin{tabular}{|c|c|c|c|c|} \hline
$\beta$&Lattice&$L_{10}=s_0^{02}$&$L_{11}=s_1^{02}$
               &$L_{12}=s_4^{02}$\\ \hline
2.3  & $ 8^4$ & 1.846~~(22)& 1.844~~(22)& 1.843 (22)  \\ 
2.3  & $12^4$ & 1.8241 (74)& 1.8217 (72)& 1.743 (12)  \\ 
2.3  & $16^4$ & 1.8307 (39)& 1.8282 (39)& 1.728 (10)  \\ \hline
2.43 & $12^4$ & 2.769~~(29)& 2.759~~(29)& 2.669 (32)  \\ 
2.43 & $16^4$ & 2.725~~(14)& 2.715~~(14)& 2.572 (18)  \\ 
2.43 & $24^4$ & 2.7395 (57)& 2.7287 (57)& 2.561 (14)  \\ 
2.43 & $28^4$ & 2.7317 (43)& 2.7212 (42)& 2.565 (12)  \\ \hline
2.51 & $16^4$ & 3.531~~(30)& 3.516~~(30)& 3.370 (31)  \\ 
2.51 & $20^4$ & 3.571~~(23)& 3.555~~(23)& 3.359 (27)  \\ 
2.51 & $28^4$ & 3.552~~(10)& 3.5371~(99)& 3.315 (18)  \\ \hline
2.574& $20^4$ & 4.356~~(37)& 4.337~~(37)& 4.084 (38)  \\ 
2.574& $24^4$ & 4.352~~(24)& 4.333~~(24)& 4.080 (29)  \\ 
2.574& $32^4$ & 4.374~~(14)& 4.355~~(14)& 4.100 (21)  \\ 
2.574& $40^4$ & 4.377~~(11)& 4.358~~(10)& 4.074 (20)  \\ \hline
2.62 & $24^4$ & 5.157~~(40)& 5.133~~(39)& 4.836 (44)  \\ 
2.62 &$24^348$& 5.070~~(30)& 5.047~~(29)& 4.788 (34)  \\ 
2.62 & $28^4$ & 5.059~~(28)& 5.037~~(28)& 4.751 (30)  \\ 
2.62 & $40^4$ & 5.068~~(15)& 5.045~~(15)& 4.725 (26)  \\ \hline
2.67 & $28^4$ & 6.021~~(46)& 5.993~~(46)& 5.603 (58)  \\ 
2.67 & $32^4$ & 5.950~~(38)& 5.923~~(38)& 5.532 (42)  \\ 
2.67 & $40^4$ & 5.910~~(25)& 5.884~~(25)& 5.536 (33)  \\ \hline
2.71 & $32^4$ & 6.656~~(51)& 6.626~~(51)& 6.208 (55)  \\ 
2.71 & $36^4$ & 6.724~~(48)& 6.692~~(48)& 6.223 (58)  \\ 
2.71 & $40^4$ & 6.656~~(31)& 6.626~~(30)& 6.188 (38)  \\ \hline
2.751&$32^364$& 7.515~~(49)& 7.481~~(48)& 7.010 (52)  \\ 
2.751& $36^4$ & 7.531~~(53)& 7.497~~(53)& 7.033 (66)  \\ 
2.751& $40^4$ & 7.576~~(46)& 7.541~~(46)& 7.038 (54)  \\ \hline
2.816& $44^4$ & 9.056~~(65)& 9.015~~(64)& 8.349 (73)  \\ \hline
2.875& $52^4$ &10.879~~(87)&10.830~~(86)&10.122 (92) \\ \hline
\end{tabular} 
\end{table} 

The cooling method was introduced in Ref.~\cite{B81} in the context of
 investigating the topological charge of the 2D O(3) sigma model. It 
has since then found many applications. For a review see \cite{N99}. 
A SU(2) cooling update maps a link matrix
\begin{eqnarray} \label{cupdt}
  U_{\mu}(n)\ \to\ U'_{\mu}(n)\,, 
\end{eqnarray}
so that $U'_{\mu}(n)$ maximizes the local contribution to the action.
This is achieved by
\begin{eqnarray} \label{cup}
  U'_{\mu}(n)\ =\ U^{\sqcup}_{\mu}(n)/{\rm det}|U^{\sqcup}_{\mu}(n)|\,,
\end{eqnarray}
where $U^{\sqcup}_{\mu}(n)$ is the staple matrix (\ref{Us}), which
for SU(2) agrees up to the determinant factor with a SU(2) matrix.

Our cooling sweeps are performed in the same systematic order as our 
MCMC sweeps. Bonati and D'Elia \cite{BD14} outlined that $n_c$ cooling 
sweeps correspond to a gradient flow time
\begin{eqnarray} \label{tc}
  t_c\ =\ n_c/3\,.
\end{eqnarray}
As we use $\epsilon=0.01$ in our gradient flow steps, one cooling 
sweep corresponds to $33.\overline{3}$ gradient sweeps. On top of 
this (because of the Runge-Kutta), one gradient sweep is more CPU time 
demanding than one cooling sweep, so that the computational efficiency 
is improved by at least a factor of 34. A priori it is not obvious that 
many small gradient flow steps can be replaced by a large cooling step 
without losing accuracy of scale setting. A posteriori our results 
support that such a replacement is permissible.

Figure~\ref{fig_t0cool} is the analogue of Fig.~\ref{fig_t40}. Due 
to the large cooling steps, gaps between them are clearly visible. 
They also exist in Fig.~\ref{fig_t40}, but are there too small to be 
noticeable. Using linear interpolations, the crossing points of the 
ratio functions (\ref{ri}) determine initial scales for the cooling 
flow in precisely the same way as explained for the gradient flow. 
The values are summarized by the equations $t^{01}=1.80$, $t^{02}=3.40$, 
\begin{eqnarray} \label{cyi01}
  y^{01}_0&=&0.0440\,,~~y^{01}_1\ =\ 0.0430\,,~~y^{01}_4\ =\ 0.0350\,,
  ~~~~\\ \label{cyi02} 
  y^{02}_0&=&0.0822\,,~~y^{02}_1\ =\ 0.0812\,,~~y^{02}_4\ =\ 0.0656\,.
  ~~~~
\end{eqnarray}
The cooling scale ${s}_i^{0j}(\beta)$ values (\ref{s0scale}) are 
collected in Tables~\ref{tab_ct01} and~\ref{tab_ct02} for the same 
lattices and $\beta$ values as used for the gradient flow. For the 
analysis in the next section these length scales are labeled by $L_7$ 
to $L_{12}$. A detailed comparison of the scaling behavior of the 
deconfinement, gradient and cooling scales follows in the next section.

\section{Scaling and asymptotic scaling} \label{sec_scaling}

In this section we analyze scaling and asymptotic scaling for 13 length 
scales
\begin{eqnarray} \label{Lk}
  L_k\,,~~~(k=0,\dots,12)\,
\end{eqnarray}
defined as follows: The deconfining scale $L_0=N_{\tau}(\beta_c)$ 
(\ref{Ntau}), six gradient, $L_1,\dots,L_6$, and six cooling, $L_7,
\dots,L_{12}$, length scales. First, we consider ${\cal O}(a^2)$
scaling corrections for length ratios in the usual way (e.g., 
\cite{L10}). Then, we incorporate asymptotic scaling behavior along 
the lines of Ref.~\cite{A97,B15} and show how this can be done in a 
way consistent with ${\cal O}(a^2)$ scaling corrections.

\subsection{Scaling} \label{sub_scaling}

\begin{table}[th] 
\centering              
\caption{\label{tab_rij}{Estimates of $r_{ij}$ ratios defined
by Eq.~(\ref{rij}).}} \smallskip
\begin{tabular}{|c|c|c|c|c|} \hline
$i\,\backslash\, j$& $L_1$ & $L_4$  & $L_7$ & $L_{10}$ \\ \hline
$L_0$   & 2.8896 (71) & 2.2290 (46) & 2.8855 (68) & 2.2618 (42) \\ \hline
$L_1$   & $-$         & 0.77382 (61)& 0.99845 (38)& 0.78433 (43)\\ \hline
$L_3$   & 0.9250 (19) & 0.7163 (17) & 0.9241 (19) & 0.7264 (16) \\ \hline
$L_4$   & 1.2943 (11) & $-$         & 1.29135 (99) & 1.01520 (49)\\ \hline
$L_6$   & 1.2090 (26) & 0.9346 (20) & 1.2081 (27) & 0.9490 (21) \\ \hline
$L_7$   & 1.00156 (38)& 0.77398 (79)& $-$         & 0.78570 (50)\\ \hline
$L_9$   & 0.9222 (21) & 0.7141 (19) & 0.9213 (20) & 0.7243 (17) \\ \hline
$L_{10}$& 1.27509 (70)& 0.98508 (47)& 1.27300 (80)& $-$         \\ \hline
$L_{12}$& 1.1835 (24) & 0.9164 (21) & 1.1825 (24) & 0.9292 (19) \\ \hline
\end{tabular} \end{table} 

To compare mass or length scales it is customary to fit ratios to the 
linear form
\begin{eqnarray} \label{rijk}
  R_{ij} = \frac{L_i}{L_j} = r_{ijk} + c_{ijk}\,\left(\frac{a}{l_k}
  \right)^2,~~l_k=a\,L_k\,,
\end{eqnarray}
where $a$ is the lattice spacing, $l_k$ the length scale in physical
units and $r_{ijk}$, $c_{ijk}$ are fit parameters of which the $r_{ijk}$ 
estimate the continuum limits and $c_{ijk}$ the leading order 
corrections. We report in Table~\ref{tab_rij} continuum estimates 
$r_{ij}$ for the subset
\begin{eqnarray} \label{rij}
  R_{ij} = r_{ij}\!+\!c_{ij}\,\left(\frac{a}{l_j}\right)^2
         = r_{ij}\!+\!c_{ij}\,\left(\frac{1}{L_j}\right)^2
\end{eqnarray}
with $i=0,1,3,4,6,7,9,10,12$ and $j=1,4,7,10$. For $i\ge 1$ gradient
and cooling scale fits, we use at each $\beta$ value our largest 
lattice and $R_{ij}$ error bars that rely on jackknife binning. In 
the case of the $L_0$ deconfinement scale, error propagation is used, 
where the values of the gradient and cooling scales at the $\beta_c$ 
values are obtained by interpolating via an asymptotic scaling fit 
performed in the next subsection.

The scales $L_2$, $L_5$, $L_8$ and $L_{11}$ are omitted from the table,
because they rely on the $E_1$ energy definition, which agrees for
practical purposes with $E_0$. For instance, $r_{11,10}=0.995397\,(24)$, 
where the error bar is very small due to correlations between the $E_0$ 
and $E_1$ energy densities. Data points from $\beta=2.3$ are eliminated 
from the fits for $q$ values smaller than $0.05$. After applying this
cut, $q$ was in the range 0.11 to 0.98.

\begin{figure}[th]\begin{center} 
\epsfig{figure=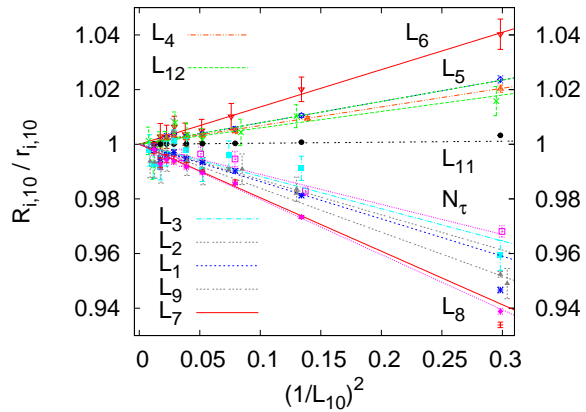,width=\columnwidth} 
\caption{Scaling corrections of order $a^2$ for ratios $L_i/L_{10}$. 
Here and in the next figures some data are slightly shifted for better 
visibility. To label all fits, some labels are attached to the lines 
and others put into the legend. The up-down order in the legend mirrors 
the up-down order in the plot. \label{fig_LiLj}} 
\end{center} \vspace{-4mm} \end{figure}

\begin{figure}[th]\begin{center} 
\epsfig{figure=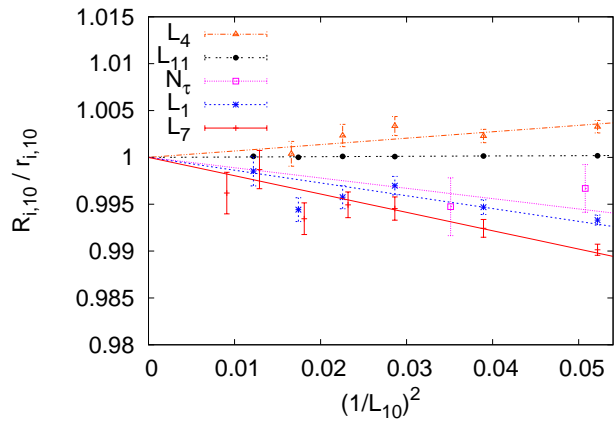,width=\columnwidth} 
\caption{Enlargement of the continuum approach of Fig.~\ref{fig_LiLj}
for the $E_0$, the $L_{11}$ and the deconfinement scales. 
\label{fig_LiLj1}} 
\end{center} \vspace{-4mm} \end{figure}

\begin{figure}[th]\begin{center} 
\epsfig{figure=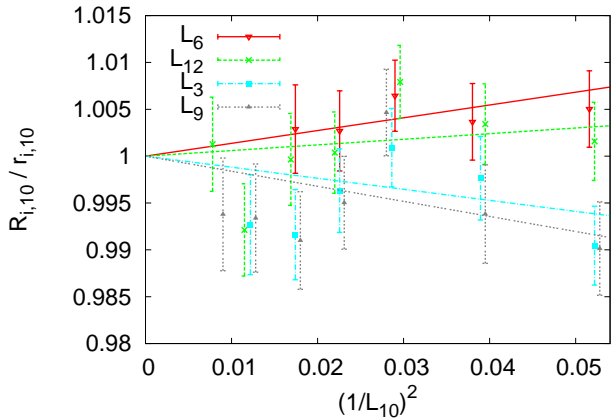,width=\columnwidth} 
\caption{Enlargement of the continuum approach of Fig.~\ref{fig_LiLj}
for the $E_4$ scales. \label{fig_LiLj2}} 
\end{center} \vspace{-4mm} \end{figure}

To compare scaling corrections we divide the $R_{ij}$ data by their 
continuum limits $r_{ij}$ and choose as reference scale $j=10$ by
reasons to be explained. A selection of the thus resulting fits is 
plotted in Fig.~\ref{fig_LiLj}. 

The fit for the deconfinement scale $N_{\tau}$ relies on the five 
$\beta_c$ data points of Tables~\ref{tab_Tc1}, \ref{tab_Tc2} and has 
a goodness of fit $q=0.25$. The $q<0.05$ cut was applied to the fits 
involving $L_{11}$, $L_2$, $L_1$ and $L_7$. For them deviations of the 
$\beta =2.3$ data points from the fit lines are clearly visible in 
Fig.~\ref{fig_LiLj} at $(1/L_{10})^2\approx 0.3$. The remaining seven 
fits include their $\beta=2.3$ data points. 

Essentially, the $L_{11}/L_{10}$ fit takes on the constant value~1. 
Similarly, $E_0$, $E_1$ pairs stay together for the other scales. 
Generally, we notice that gradient and cooling scales that use the 
same energy observable and target value $y_i^{01}$ or $y_i^{02}$ stay
closer together than gradient scales using different energy observables 
and target values or cooling scales using different energy observables 
and target values. The ratios of Table~\ref{tab_rij} show the same 
pattern. So, it appears perfectly legitimate to use cooling instead 
of gradient scales. We opted for $L_{10}$ as reference scale, because 
it centers rather nicely with respect to the other scales. At 
$(1/L_{10})^2\approx 0.3$ in Fig.~\ref{fig_LiLj} we read off scaling 
violations of about 10\%, i.e., 0.94 to 1.04 for $R_{i,10}/r_{i,10}$. 
That is larger than the 5\% reported by L\"uscher \cite{L10} in his 
Fig.~3 for SU(3) at $\beta=5.96$. As outlined, this corresponds to 
$\beta\approx 2.46$ for SU(2), which translates into $(1/L_{10})^2
\approx 0.11$. In Fig.~\ref{fig_LiLj} this is slightly left of the 
column of data at $(1/L_{10})^2\approx 0.13$ for which we find the 
range $0.97<R_{i,10}/r_{i,10}<1.02$, i.e., scaling violations are 
down to less than~5\%.

A problem with plots like Fig.~\ref{fig_LiLj} is that data from large 
lattices (close to the continuum limit) accumulate in a small region, 
which is here below $(1/L_{10})^2<0.05$. It is enlarged in 
Figs.~\ref{fig_LiLj1} and~\ref{fig_LiLj2}. In Fig.~\ref{fig_LiLj2} 
the length scales based on the $E_4$ energy are plotted and seen to 
exhibit considerably larger error bars than the energy scales plotted 
in Fig.~\ref{fig_LiLj1}. With no particular advantages to offset this 
lack of accuracy of the $E_4$ scales, all arguments converge in favor 
of using an $E_0$ cooling scale.

\subsection{Asymptotic scaling} \label{sub_ascaling}

For large $\beta$ the scaling of any mass $m$ in pure SU(N) LGT is 
determined by the asymptotic scaling function
\begin{eqnarray} \nonumber
  &~& a\,m\ =\ {\rm const}\,f_{as}(\beta)\,, \\ \nonumber
  &~& f_{as}(\beta)\ =\ \alpha\,a\Lambda_L=\alpha\,\left(b_0\,\frac{2N}{\beta}
  \right)^{-b_1/2b^2_0}\\ \label{fas} &\times& \exp\left(-\frac{
  \beta}{4N\,b_0}\right)\,\left[1+\sum_{j=1}^{\infty} q_j\,
  \left(\frac{2N}{\beta}\right)^j\right]\,,
\end{eqnarray}
where $a$ is the lattice spacing, $b_0=11\,N/(48\pi^2)$ and $b_1=
(34/3)\,N^2/ (16\pi^2)^2$ are, respectively, the universal 1-loop 
\cite{Gr73,Po73} and 2-loop \cite{Jo74,Ca74} asymptotic scaling 
coefficients. Universal means that all renormalization schemes lead to 
the same $b_0$ and $b_1$ values. Non-universal perturbative corrections 
are given by the $q_j$ coefficients in the bracket. Computing up to 
3-loops, All\'es et al.\ \cite{AF97} calculated $q_1$ for SU(N) LGT
and
\begin{eqnarray} \label{q1}
  q_1\ =\ 0.08324~~{\rm for\ SU(2)}\,.
\end{eqnarray}
Further, we introduce the factor $\alpha$ to enforce for SU(2) the 
convenient normalization
\begin{eqnarray} \label{b2}
   f_{as}(2.3)\ =\ 1\,.
\end{eqnarray}
Higher orders corrections in the lattice spacing $a$ are reflected 
by terms of the form 
\begin{eqnarray} 
  (\alpha\,a\Lambda_L)^i = [f_{\rm as}(\beta)]^i\,,~~(i=2,3,\dots)\ . 
\end{eqnarray}
Following Allton \cite{A97} in the version of \cite{B15} we arrive 
at the expansions
\begin{eqnarray} \label{Lkas} 
  L_k = \frac{c_k}{f_{as}(\beta)}\,\left(1+\sum_{i=1}^{\infty}
       a_k^i\,[f_{as}(\beta)]^i\right)
\end{eqnarray}
for our length scales, where $c_k$ and the $a^i_k$ are parameters that
have to be calculated. In practice we have to truncate the series 
(\ref{Lkas}) as well as the definition (\ref{fas}) of $f_{as}(\beta)$.
Defining
\begin{eqnarray} \label{f0as} 
  f^0_{as}(\beta)&=&\alpha^0\,\left(b_0\,\frac{2N}{\beta}\right)^{-b_1
  /2b^2_0}\exp\left(-\frac{\beta}{4N\,b_0}\right)\,,~~~\\ \label{f1as}
  f^1_{as}(\beta)&=& \left(\frac{\alpha^1}{\alpha^0}\right)\,
  f^0_{as}(\beta)\,\left(1+\frac{4\,q_1}{\beta}\right)\,,
\end{eqnarray}
we have $f^m_{as}$ with $m=0,1$ at our disposal, where the coefficients
$\alpha^m$ are defined to enforce as in (\ref{b2}) the normalizations 
$f^m_{as}(2.3)=1$. Truncating the sum (\ref{Lkas}) by fixed $n$, we end 
up with 26 fits ($k=0,\dots,12$), ($m=0,1$):
\begin{eqnarray} \label{Lkmn} 
  L_k^{m,n} = \frac{c_k^{m,n}}{f^m_{as}(\beta)}\,\left(1+\sum_{i=1}^n
       a_k^{m,i}\,[f^m_{as}(\beta)]^i\right)\,,
\end{eqnarray}
where the index $n$ of $a_k^{m,i}$ is suppressed. Due to the truncation 
of $f_{as}$ there are perturbative corrections in $1/\beta$ when ratios 
are taken with respect to the (inverse) lambda lattice scale, i.e.,
\begin{eqnarray} \label{Rlambda} 
  L_k^{m,n}\,\alpha^m\,a\Lambda_L = c_k^{m,n} + 
  {\rm perturbative\ corrections}
\end{eqnarray}
describes asymptotic scaling. Corrections to ratios of two length scales 
are exponentially small in $\beta$, i.e., 
\begin{eqnarray} \label{Rkmn} 
  \frac{L_{k_1}^{m,n_1}}{L_{k_2}^{m,n_2}} = 
  \frac{c_{k_1}^{m,n_1}}{c_{k_2}^{m,n_2}} + 
  {\rm non\!\!-\!\!perturbative\ corrections}\
\end{eqnarray}
holds. However, due to the $a^{m,1}_k$ term in (\ref{Lkmn}) corrections 
would in general be of order $a$ in the lattice spacing and not of order 
$a^2$ as in (\ref{rij}). In \cite{B15} this problem was avoided by 
combining several scales into one fit. This is only possible when 
their relative scaling violations are so weak that they become invisible 
within statistical errors. The solution which we propose here is to fit 
all $k=0,1,\dots,12$ scales with identical $a^{m,1}_k$ coefficients so
that the non-perturbative corrections (\ref{Rkmn}) become ${\cal O}(a^2)$.

\begin{table}[th] 
\centering 
\caption{\label{tab_as}{Asymptotic scaling fits of normalization 
constants and goodness of fit $q$.  }}
\smallskip
\begin{tabular}{|c|c|c|c|c|c|c|} \hline
$k$& $c_k^{1,3}$&     & $c_k^{0,4}$& $q$ & $c_k^{1,4}$& $q$ \\ \hline
 0 & 6.6682 (56)& 0.00& 6.114 (29) & 0.71& 5.892 (27) & 0.68\\ \hline
   & $c_k^{1,2}$&     &$c_k^{0,3}$ &     & $c_k^{1,3}$&     \\ \hline
 1 & 2.2481 (32)& 0.04& 2.1937 (64)& 0.91& 2.1083 (61)&0.91\\ 
 2 & 2.2311 (32)& 0.03& 2.1812 (64)& 0.92& 2.0961 (60)&0.92\\ 
 3 & 2.0743 (56)& 0.17& 2.022~~(11)& 0.66& 1.9432 (98)&0.67\\ 
 4 & 2.8945 (54)& 0.08& 2.846~~(11)& 0.98& 2.735~~(11)&0.98\\ 
 5 & 2.8835 (53)& 0.04& 2.837~~(11)& 0.98& 2.727~~(11)&0.98\\ 
 6 & 2.7068 (85)& 0.95& 2.658~~(18)& 0.95& 2.555~~(17)&0.95\\ 
 7 & 2.2498 (30)& 0.02& 2.1996 (61)& 0.93& 2.1138 (57)&0.94\\ 
 8 & 2.2254 (30)& 0.01& 2.1807 (60)& 0.92& 2.0956 (57)&0.93\\ 
 9 & 2.0664 (58)& 0.16& 2.018~~(11)& 0.69& 1.9397 (99)&0.69\\ 
10 & 2.8501 (46)& 0.02& 2.8037 (91)& 0.89& 2.6942 (86)&0.89\\ 
11 & 2.8357 (45)& 0.01& 2.7914 (89)& 0.88& 2.6824 (85)&0.89\\ 
12 & 2.6485 (74)& 0.26& 2.599~~(14)& 0.52& 2.498~~(13)&0.52\\ \hline
   &\multicolumn{2}{c|}{$a_k^{1,1}=-0.6209$}
   &\multicolumn{2}{c|}{$a_k^{0,1}=-0.38157$}
   &\multicolumn{2}{c|}{$a_k^{1,1}=-0.32536$} \\ \hline
\end{tabular} 
\end{table} 

Estimates of normalization constants for asymptotic scaling fits are 
collected in Table~\ref{tab_as}. As before, the gradient and cooling 
scale fits use our largest lattice at each $\beta$ value. The last row 
of the table gives the $a_k^{m,1}$ values taken 
for all fits of their respective columns. Using the $E_0$ and $E_4$ 
scales these values were determined by the maximum likelihood method 
($E_1$ scales are left out because they would in essence amplify weights
of the $E_0$ scales). On a technical note, we remark that we eliminate 
the normalization constants $c_k^{m,n}$ from the search for the $\chi^2$ 
minimum by treating them as functions of the $a_k^{m,i}$ parameters 
\cite{B16}. This stabilizes the search considerably, for which we used 
the Levenberg-Marquardt approach (e.g., \cite{B04}).

Fitting the gradient and cooling scales ($k\ge 1$) with only one 
additional parameter, $a_k^{1,2}$, the normalization constants 
$c_k^{1,2}$ of column two are obtained. Most $q$ values of these 
fits are too low. So, we decided to allow for one more fit parameter, 
$a_k^{m,3}$. The results are shown in columns four and six (using 
$f_{as}^m$ with $m=0,1$). Now, the $q$ values for these fits would be 
too good to be true if they were statistically independent. As they all 
rely on the same data set correlations can explain that a whole series 
of fits exhibits $q>0.5$, mostly close to 0.9. Notably, consistent fits 
due to adding the $a_k^{m,3}$ parameters come at the price of about 
doubled error bars compared to those of column two.

It is possible to include the deconfinement length scale into these fits 
with fixed $a_k^{m,1}$ and the results are given in the second row of 
Table~\ref{tab_as}. Despite the small number of only five data points 
(Tables~\ref{tab_Tc1} and \ref{tab_Tc2}) one needs one more parameter, 
$a_0^{m,4}$, to get acceptable $q$ values. This is accompanied by some
instability discussed at the end of this section. 

Using the $f^1_{as}$ instead of the $f^0_{as}$ asymptotic scaling 
function decreases all $c_k$ values by slightly less than 4\%. More 
prominent is the decrease between 6.7\% to 9\% from column two to 
column six, which comes from allowing one more free parameter. Together 
we take this as an indication that remaining systematic errors may well 
reach~10\%.

\begin{figure}[th]\begin{center} 
\epsfig{figure=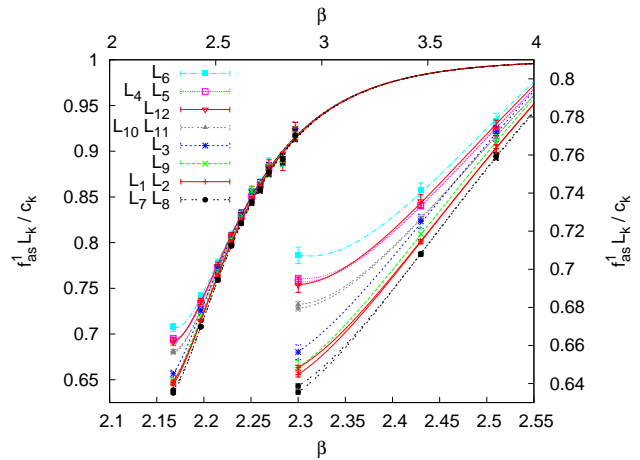,width=\columnwidth} 
\caption{Asymptotic scaling.  \label{fig_ascaling}} 
\end{center} \vspace{-4mm} \end{figure}

Dividing out the asymptotic scaling behavior $c_k^{1,n}/f^1_{as}(\beta)$,
we plot in Fig.~\ref{fig_ascaling} the resulting fits $f^1_{as}
\,L_k^{1,3}/c_k^{1,3}$ ($k\ge 1$) for column six of Table~\ref{tab_as}. 
For the curves on the left the abscissa is on top of the figure and the 
ordinate on the left. At $\beta=4$ all fits have almost reached the 
asymptotic value~1. The lower abscissa and the right ordinate apply 
to the right part of Fig.~\ref{fig_ascaling}, which enlarges the range 
of our initial three $\beta$ values. At $\beta=2.3$ asymptotic scaling 
violations are seen to range from 28\% to 37\%. The relative differences 
reach only $0.72/0.63\approx 1.14$, consistent with the ratio $1.04/0.93
\approx 1.12$ observed at $(1/L_{10})^2=0.3$ in Fig.~\ref{fig_LiLj}.

Let us turn to the scaling behavior of ratios. Except for the 
deconfinement length scale $L_0$, which is statistically independent 
from the other scales, we cannot use error propagation. Instead, we 
calculate the $R_{ij}$ ratios (\ref{rij}) for jackknife bins built 
from the individual gradient or cooling flow runs (using jackknife bins 
of the asymptotic scaling fits of Table~\ref{tab_as} has the problem 
that these fits have larger fluctuations than the $R_{ij}$ ratios).

\begin{table}[th] 
\centering        
\caption{\label{tab_ratb}{Estimates of $r_{ij}$ ratios from scaling 
fits of jackknifed $R_{ij}$ data.}} \smallskip
\begin{tabular}{|c|c|c|c|c|} \hline
$i\,\backslash\,j$& $L_1$& $L_4$    & $L_7$       & $L_{10}$    \\ \hline
$L_0$ (as)& 2.795 (16)& 2.154  (14) & 2.787  (15) & 2.187  (13) \\ \hline
$L_0$   &*2.914  (15) & 2.2393 (52) &*2.903  (14) & 2.2692 (48) \\ \hline
$L_1$   & $-$         &*0.7703 (12) & 0.99808 (34)&*0.78185 (77)\\ \hline
$L_3$   & 0.9240 (20) & 0.7187 (19) & 0.9221 (20) & 0.7275 (17) \\ \hline
$L_4$   &*1.2996 (21) & $-$         &*1.2957 (27) & 1.01373 (57) \\ \hline
$L_6$   & 1.2000 (31) & 0.9334 (23) & 1.1972 (32) & 0.9465 (24) \\ \hline
$L_7$   & 1.00188 (34)&*0.7728 (16) & $-$         &*0.78419 (88)\\ \hline
$L_9$   & 0.9214 (22) & 0.7171 (21) & 0.9197 (22) & 0.7255 (18) \\ \hline
$L_{10}$&*1.2795 (13)& 0.98638 (55) &*1.2760 (15) & $-$         \\ \hline
$L_{12}$& 1.1786 (26) & 0.9167 (24) & 1.1760 (26) & 0.9283 (20) \\ \hline
\end{tabular} \end{table} 

For $m=1$ results are collected in Table~\ref{tab_ratb}. With 
exception of the $L_0$ (as) row (to be discussed) all fits use
\begin{eqnarray} \label{ak11} 
   a_k^{1,1}\ =\ 0 
\end{eqnarray}
to reflect that the leading scaling corrections for mass ratios 
are ${\cal O}(a^2)$. We end up with
\begin{eqnarray} \label{Rijas} 
  R_{ij}\ =\ r_{ij}+\sum_{i=2}^n a^{1,i}\left[f^1_{as}\right]^i
\end{eqnarray}
Surprisingly, one additional free parameter $a_k^{1,2}$, besides the 
ratio estimate $r_{ij}$, gives in more than half of the cases a 
satisfying goodness of fit ($0.13\le q\le 0.99$). For the other cases, 
indicated by * in Table~\ref{tab_ratb}, the parameter $a_k^{1,3}$ is 
also needed ($0.45\le q\le 0.75$ holds for these). Comparing with 
our previous ratio estimates of Table~\ref{tab_rij}, we see that the 
error bars of the starred estimates are about two times larger, while 
the error bars of the other estimates are similar as before. Systematic 
errors due to the different fits are around 1\%, which is up to an 
order of magnitude larger than the statistical errors. The latter can 
be extremely small due to correlations between the estimators.

Using the asymptotic scaling function with $m=0$ instead of $m=1$,
differences for ratios are about two orders of magnitude smaller than 
those encountered for the normalization constants of Table~\ref{tab_as}. 
Asymptotic scaling corrections drop out, as one expects. The systematic 
error due to adding the $a_k^{m,3}$ fit parameter can be considerably 
larger, up to 1.3\%. This is still about one magnitude smaller than the 
same systematic uncertainty in the case of the normalization constants.

\begin{figure}[th]\begin{center} 
\epsfig{figure=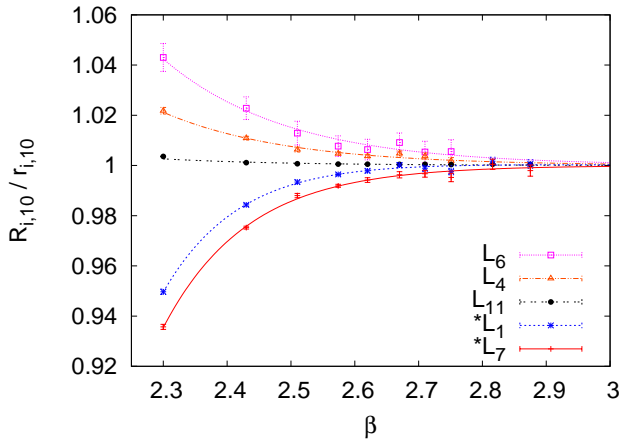,width=\columnwidth} %
\caption{Scaling corrections of the $E_0$, the $L_{11}$ and the 
deconfinement scale ratios with respect to $L_{10}$. \label{fig_scaling1}} 
\end{center} \vspace{-4mm} \end{figure}

\begin{figure}[th]\begin{center} 
\epsfig{figure=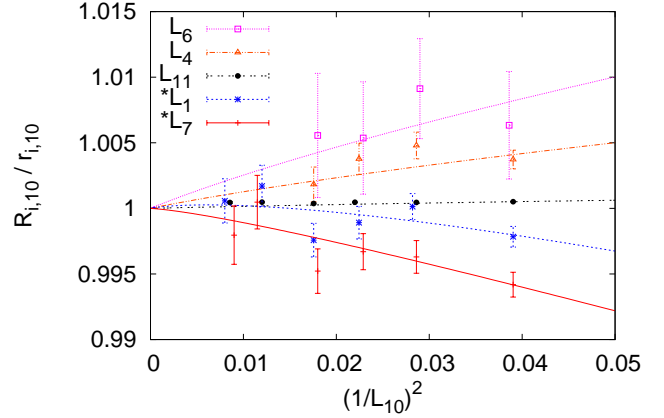,width=\columnwidth} 
\caption{Figure~(\ref{fig_scaling1}) plotted in the $(1/L_{10})^2$ 
range of Fig.~\ref{fig_LiLj1}. \label{fig_scaling2}} 
\end{center} \vspace{-4mm} \end{figure}

Dividing the constants $r_{ij}$ out, Figs.~\ref{fig_scaling1} 
and~\ref{fig_scaling2} give a visual impression of the scaling of
selected fitting curves with reference scale $L_{10}$. Superficially,
curves for the same scales look similar in Fig.~\ref{fig_scaling2}
as before in Figs.~\ref{fig_LiLj1} and~\ref{fig_LiLj2}. However, there 
is a fundamental difference between the fits. Equation (\ref{Rijas}) 
ensures that $L_i/L_{10}\sim (1/L_{10})^2$ is correct for in the limit
$(1/L_{10})^2\to 0$, while in Eq.~(\ref{rij}) it is assumed to be 
already exact for the data at hand. Now, for the fits (\ref{Rijas}) 
the straight line behavior is in some cases only reached for very small 
$(1/L_{10})^2$. This is most pronounced for the $R_{1,10}/r_{1,10}$ 
fit, which crosses the value~1 from below and finally approaches~1 from 
above, once the region $(1/L_{10})^2<0.005$ on the very left side of 
Fig.~\ref{fig_scaling2} is reached (details are not visible on the scale
of the figure). In view of this it is reassuring that the estimates of 
Tables~\ref{tab_rij} and~\ref{tab_ratb} never differ by more than 1.3\%.
The two fitting approaches supplement one another and give some insight
into systematic errors one may expect.

\begin{figure}[th]\begin{center} 
\epsfig{figure=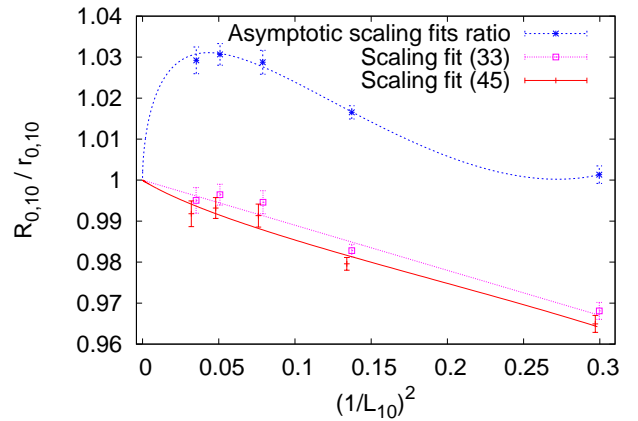,width=\columnwidth} 
\caption{Three fits of the deconfinement length scale $L_0$ versus 
$(1/L_{10})^2$. \label{fig_Tc10}} 
\end{center} \vspace{-4mm} \end{figure}

We conclude this section discussing the instabilities encountered 
when fitting $L_0/L_i$. In the $L_0$~(as) row of 
Table~\ref{tab_ratb} we report estimates obtained from using the 
constants of column six of Table~\ref{tab_as} and error propagation. 
Compared with the previous estimates of Table~\ref{tab_rij} we find 
a systematic decrease in the range 3.2\% to 3.6\%, larger than the 
statistical error, which never exceeds 0.6\%. As the asymptotic 
scaling of $L_0$ needs four parameters to fit just five data points 
one may suspect ``overfitting''. As a tiebreaker we perform the fit 
of Eq.~(\ref{Rijas}) for jackknifed ratios of $L_0/L_j$, $j=1,4,7,10$ 
and obtain the estimates of the $L_0$ row of Table~\ref{tab_ratb}. The 
systematic errors with respect to Table~\ref{tab_rij} are now down to 
less than~1\%.

Dividing the asymptotic ratios out, the three fits for $L_0/L_{10}$ 
are shown in Fig.~\ref{fig_Tc10}. The straight line fit from 
Figs.~\ref{fig_LiLj} and~\ref{fig_LiLj1} comes in as second lowest. 
The lowest curve corresponds to Eq.~(\ref{Rijas}) and the upper curve 
to dividing the $L_0$ fit of column six of Table~\ref{tab_as} by the 
$L_{10}$ fit of the same column. As suspected this curve looks rather 
fanciful. However, using a log scale for the abscissa would stretch
the range on the left, and one should have in mind that the absolute 
differences between all three fits are quite small. Systematic errors 
at $(1/L_{10})^2=0.3$ can be read off on the right-hand side of the 
figure and are seen to be less than~4\%.

\section{Summary and conclusions} \label{sec_sum}

We have studied the approach of SU(2) LGT to its quantum continuum limit 
by investigating the scaling behavior of a number of length scales with
definitions based on the deconfinement phase transition, the gradient 
flow and the cooling flow. While the deconfining scale $L_0=N_{\tau}$ 
is uniquely defined (\ref{Ntau}), one has considerable freedom in the 
definition of gradient and cooling flow scales. They depend on the 
choice of observables and target values of the flow. We considered:

\begin{enumerate}
\item Energy densities $E_0$, $E_1$, $E_4$ defined by Eqs.~(\ref{E0}, 
\ref{E1}, \ref{E4}). $E_0$ is up to normalization the Wilson action 
and $E_1$ in essence an equivalent definition. $E_4$, introduced in 
\cite{L10}, averages over four plaquettes.
\item Target values $y^{01}_i$ and $y^{02}_i$, ($i=0,1,4$) are defined 
by Eqs.~(\ref{yi01}, \ref{yi02}, \ref{cyi01}, \ref{cyi02}). They are 
constructed so that the initial scaling behavior of either the gradient 
or the cooling flow of either $E_0$, $E_1$ or $E_4$ matches that of the 
deconfinement length $N_{\tau}$ (altogether $3\times 4=12$ distinct 
definitions).
\end{enumerate}

For ratios of these length scales, corrections to scaling are supposed 
to be of order $a^2$ in the lattice spacing as illustrated in 
Figs.~\ref{fig_LiLj}, \ref{fig_LiLj1}, \ref{fig_LiLj2}, 
\ref{fig_scaling2} and~\ref{fig_Tc10}. In these figures the 
cooling length scale $L_{10}$, which relies on the $E_0$ energy
density and a $y^{02}_0$ target value (\ref{cyi02}), is used 
as reference scale by the following reasons:

\begin{enumerate}
\item Scaling violations of ratios of scales are then rather symmetrically 
distributed above and below~1.
\item $E_0$ is easier to calculate than $E_4$ and estimates from the 
same statistics result in smaller error bars for the $E_0$ length scale. 
No scaling advantages were found for $E_4$ scales. $E_1$ is essentially 
equivalent to $E_0$ with the benefit for $E_0$ that the Wilson action 
is implemented in the program anyhow.
\item The cooling flow is faster and easier to calculate than
the gradient flow and there is no noticeable loss of accuracy
as anticipated in Ref.~\cite{BD14}. As the cooling method \cite{B81} 
was an answer to difficulties encountered when trying to calculate
the topological charge in a paper by L\"uscher and one of the authors 
\cite{BL81}, it appears that the cooling scale could have been 
introduced 30 years before the gradient scale \cite{L10}. 
\end{enumerate}

The magnitude of scaling violations we find for ratios of length scales 
is close to that reported in Ref.~\cite{L10} for SU(3) when comparing 
the $E_0$ with the $E_4$ flow. The SU(2) scaling region begins at 
$\beta=2.3$ where we find corrections to scaling in the 10\% range. 
Deeper in the scaling region, at $\beta=2.46$, they become reduced 
to slightly less than~5\%.

Scaling corrections for the ratio $N_{\tau}/L_{10}$ fall into the 
range provided by the other scales as is seen in Figs.~\ref{fig_LiLj} 
and~\ref{fig_LiLj1}. The significant advantage of the gradient scale, 
and to an even greater extent the cooling scale, over the deconfinement 
scale is that we can far more easily follow the scaling behavior towards 
the continuum limit. On the other hand, there are no ambiguities in the 
definition of the deconfinement scale, which makes it kind of ideal 
to define initial scaling values as discussed in sections~\ref{sec_grad} 
and~\ref{sec_cool}.

We have used two rather different approaches for analyzing our data.
For Figs.~\ref{fig_LiLj} to~\ref{fig_LiLj2} we simply calculate $L_i/
L_{10}$ from jackknife bins of the data and perform the linear 
2-parameter fit (\ref{rij}) using the ${\cal O}(a^2)$ dependence 
$(1/L_{10})^2$ from the same data. While this is straightforward, 
one  does not connect with the asymptotic $\Lambda_L$ scale. 

To connect with asymptotic scaling, we relied on truncated forms 
of Eq.~(\ref{fas}) based on Ref.~\cite{A97,B15}. 
The normalization constants of our asymptotic scaling fits are collected 
in Table~\ref{tab_as}. A common fixed parameter ensures that scaling 
corrections for ratios are~${\cal O}(a^2)$. Systematic errors due to 
distinct truncations of the fits are found around 10\%. 
For the gradient and cooling scales the finally accepted fits of column 
six rely on three free parameters, one of them being the normalization 
constant that yields the continuum estimate. 
For $L_0$ four fit parameters are needed despite the fact that there 
are only five data points. Comparing in Fig.~\ref{fig_Tc10} the 
ratio of the $L_0$ and $L_{10}$ fit with direct fits of the $R_{0,10}$
ratios indicates overfitting, though $L_0$ data on larger lattices
is needed to be conclusive.

While the lattice spacing is exponentially small in $\beta$, asymptotic 
scaling corrections come in powers of $1/\beta$. As is seen in 
Fig.~\ref{fig_ascaling}, they range at $\beta=2.3$ from 30\% to 36\%. 
The scales cluster together, so that the relative deviations at $\beta
=2.3$ reproduce the previously encountered 10\% range. 

For ratio estimates it turns out that one should not divide the 
asymptotic scaling estimates by one another, but perform the fit 
(\ref{Rijas}) for the jackknifed $R_{ij}$ ratios of the data, where 
the common fixed parameter is set to zero to enforce ${\cal O} (a^2)$ 
corrections.  A decisive difference to the previous approach (\ref{rij}) 
remains: The ${\cal O}(a^2)$ behavior is no longer enforced for our data 
at hand, but only in the continuum limit. Indeed, some of the fits make 
use of this possibility. Compare Fig.~\ref{fig_scaling2} with 
Figs.~\ref{fig_LiLj1} and~\ref{fig_LiLj2}. Despite the differences 
in the approach to the continuum limit, the obtained curves look 
similar.

The continuum limit estimates of our ratios are collected in 
Tables~\ref{tab_rij} and~\ref{tab_ratb} using, respectively, (\ref{rij}) 
and (\ref{Rijas}). Differences due to the distinct fit forms stay below 
1.3\%. This is in most cases larger than the statistical errors. The 
different fit forms allow one to get an idea of the systematic errors 
possible.

In conclusion, we hope that the methods outlined are also of some value
for studying the approach of physically realistic theories like QCD 
to their continuum limits. Though such data rely on large scale 
calculations on supercomputers, it is presumably safe to assume 
that their quality is not better than that of our SU(2) data.

\acknowledgments
David Clarke was in part supported by the US Department of Energy 
(DOE) under contract DE-SC0010102. Our calculations used resources 
of the National Energy Research Scientific Computing Center (NERSC), 
a DOE Office of Science User Facility supported by the DOE under 
Contract DE-AC02-05CH11231.

\end{document}